# A quantitative astronomical analysis of the Orion Correlation Theory

Vincenzo Orofino

(Dipartimento di Matematica e Fisica "E. De Giorgi", Università del Salento, Lecce, Italy)

**Abstract**

The link between the three major Giza pyramids and the stars of the Orion Belt has been since long time the subject of various qualitative speculations. In this framework an important role is played by a controversial theory, the so-called "Orion Correlation Theory" (OCT), according to which a perfect coincidence would exist between the mutual positions of the three stars of the Orion Belt and those of the main Giza pyramids.

In the present paper the OCT has been subjected to some quantitative astronomical and astrophysical verifications, in order to assess its compatibility with the results of both naked-eye astrometry and photometry. In particular, a linear correlation is found between the height of such monuments and the present brightness of the Orion Belt stars. According to these analyses it is possible to conclude that the OCT is not incompatible with what expected for the stars of the Orion Belt on the basis of naked-eye astrometry and photometry, as well as of the stellar evolution theory.

# 1. Introduction: the Egyptian astronomy

According to various Egyptologists (Neugebauer, 1969, 1976; Thurston, 1994; Wells, 1996), the astronomical knowledge of Egyptians during the Old Kingdom (2700 – 2200 B.C.) was very poor. This negative opinion comes mainly from the evidence that no purely astronomical Egyptian texts dating back to that epoch have been found yet. According to Magli (2006), however, this does not mean that such texts have never been written, but it can be simply due to the fact that these papyri did not belong to the funeral kits from which up to now almost all the archaeological finds come.

The interest of the ancient Egyptians in the celestial phenomena is suggested by various inscriptions found out on the sarcophagi of the Middle Kingdom (1990 – 1780 B.C.) and in the tombs of the New Kingdom (1530 – 1080 B.C.), such as the famous tomb of Senmut, a dignitary of the queen Hatshepsut (ca. XV century B.C.). Furthermore, it is now sure that the ancient Egyptians orientated their monuments and sacred buildings toward the rising points of some bright stars, such as Sirius and Canopus, and that they used other stars (namely those of the constellation Ursa Major, or the Big Bear) to align temples and pyramids with the cardinal points (see Belmonte et al., 2008, and references therein).

As early as the Middle Kingdom, the Egyptian astronomers were able to track the movements (recording the times of rising, culmination, setting, the period of invisibility and so on) of a set of 36 stars, or small groups of stars, called "decans", mainly used for time keeping (Magli, 2006). Furthermore, since the Second Dynasty (2650 a.C.) the High Priest of the sanctuary of Heliopolis was called the "Chief of the Observers", and this testifies that during the Old Kingdom astronomical observations were surely one of the main duties of some Egyptian priests (Magli, 2009a). In fact as early as the Old Kingdom, they knew the five planets Mercury, Venus, Mars, Jupiter and Saturn, due to their movement with respect to the "fixed" stars.

For this reason, even if compelling evidence of the observation of decans exists only in the archaeological finds of the Middle Kingdom or later, it is not unreasonable to suppose that such observations could have their roots in much more ancient astronomical practices. Some authors (Pogo, 1930; Zaba, 1953; Sellers 1992; Bauval, 2006), have even suggested that Egyptians were aware of the phenomenon of equinox precession since very ancient epochs, well before its official discovery attributed to the Greek astronomer and mathematician Hipparchus (I –II century a.C.).

But the field where the Egyptian astronomers undisputedly excelled was the measure of time, in particular the evaluation of the length of the year, with the consequent introduction of the calendar, a powerful tool for agriculture that was the most important economic activity of the country. In the ancient civil Egyptian calendar the year was 365 days long and was divided into three seasons corresponding with the hydrological cycle of the Nile. Each season was divided into four months of thirty days each, making a total of 120 days in each season. At the end of the year, five extra days were added to make up the 365-day total.

In order to keep time during the daytime, the latter was divided in 12 hours of variable length, according to the season. They were determined using the motion of the Sun, by means of timepieces similar to sundials. As far as the 12 hours of night-time are concerned, the Egyptian priests-astronomers used stellar water clocks (similar to sand hourglasses) with notches indicating the hours. The notches were engraved using the motion of the above cited decans, whose rising defined the beginning of each hour. In practice the Egyptians calibrated their water clocks by means of the decan stars. The Egyptians elaborated a complex system which, each night, made use of a set of 12 decans that changed in the course of the year: every ten days the westernmost decan was dropped from the set and a new one in the east was added (Magli, 2006).

Among all the decans, Sirius ($\alpha$ Canis Majoris) was of special importance in Egypt, not only because it is the brightest star in the sky. In fact, when the calendar was adopted, the first day of the year, that coincided with the beginning of the flood of the Nile river, corresponded to the heliacal rise of Sirius. In that particular day the star, after a 70-day period of invisibility, could be observed again, rising on the horizon just before sunrise.

Other important decans were the stars of the Orion Belt, a linear asterism of three evenly spaced objects in which the northernmost star, Mintaka (or $\delta$ Orionis), is slightly out of the axis connecting the southernmost object, Alnitak (or $\zeta$ Orionis), to the central one, Alnilam (or $\epsilon$ Orionis). Even if alternative interpretations exist (Baux, 1993; Legon, 1995), it is commonly thought that the ancient Egyptians associated the Orion constellation (Sah), and in particular the Orion Belt, to Osiris, one of the most important gods of the Egyptian Pantheon, while the star Sirius (Sopdet or Sothis) represented the goddess Isis, sister and wife of Osiris (Bauval, 2006).

In any case it seems certain that the Egyptians believed in a heavenly after-life in which the souls of the dead transmigrated after their death. According to the Pyramids Texts such a heavenly kingdom was located in that region of the sky placed around the Orion Belt. In this context the pyramids were probably built to assist the pharaoh in his journey to the next life. It was believed that subjecting the dead body of the pharaoh to certain rituals and burying it in a pyramid allowed his soul to ascend to the sky, becoming a star (Badawy, 1964).

**2. Comparison between the positions of the Giza pyramids and those of Orion Belt stars**

In the contest of the above discussed link between stars and pyramids, an important role is played by a controversial theory proposed by Bauval e Gilbert (1994), the so-called "Orion Correlation Theory" (OCT). According to these authors, a perfect coincidence would exist between the mutual positions of the three stars of the Orion Belt and those of the main Giza pyramids, so that the latter would represent the monumental reproduction on the ground of that important asterism.

In the present paper the whole question is reanalyzed, subjecting the OCT to some quantitative astronomical and astrophysical verifications, in order to assess the compatibility of this theory with the results of both naked-eye astrometry and photometry.

To compare the positions of the Giza pyramids with those of the Orion Belt stars, a stellar map of that region of the sky[1] has been conveniently rescaled (scale factor of 0.003 °/m) and rotated (anti-clockwise rotation of 195.3°) and has been overlapped to the topographic map of the Giza necropolis, as shown in Fig. 1.

---

[1]: For sake of simplicity I used a modern star chart, since the proper motions of the stars of the Orion Belt are so low that their effects on the stellar coordinates at the epoch of the pyramids are negligible. In fact, using the Precession Routine of the astronomical library Starlink (available at http://fuse.pha.jhu.edu/support/tools/precess.htm) coupled with SIMBAD data of the three stars, it is possible to find that the angular separation between Alnilam and Mintaka

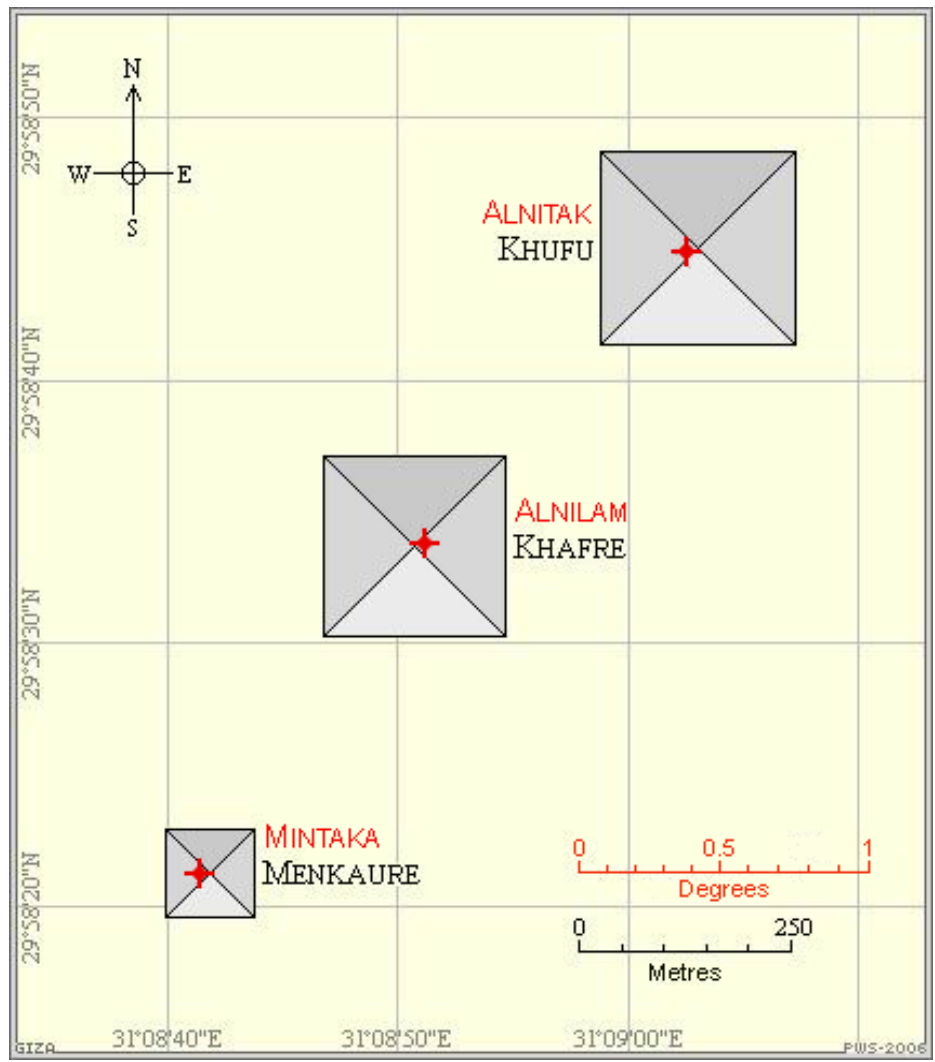

**Figure 1** – Comparison between the positions of the Orion Belt stars (red dots) and those of the vertexes of the corresponding pyramids. The star map (scale in red) has been overlapped to the topographic map of the Giza necropolis (scale in black with the related coordinate grid in gray - after Wakefield Sault, 2008). The crosses represent the error bars on the stellar positions and indicate the minimum uncertainty, equal to ±3', due to the resolution power of the human eye under optimal observation conditions (see text).

This figure shows that a certain discrepancy exists between the actual position of the vertex of each pyramid and the position expected on the basis of the stellar correlation; such a difference is more pronounced in the case of the couple Khufu-Alnitak where it is equal to 3.1% of the angular

---

decreased by only 3" from today to 2500 BC (much less that the resolution power of the human eye – see text), while the angular distance between the other two couples of stars suffered an even smaller variation.

distance between Alnitak and Alnilam; since the angular separation of the two stars is 1.356°, this corresponds to about 2.5'.

This value is less than the resolution power of the human eye (defined as the minimum angular distance between two sources necessary to see them as distinct objects); the latter in general falls between 5' and 10' (Silvestro, 1989), according to the characteristics of the observed sources, and, in the most favorable cases, can be as low as 3' (Herrmann, 1975; Gribbin e Gribbin, 1996). Any measurement of stellar position (astrometry), performed with unaided eye, can never have an uncertainty less than the resolution power of the human eye, and in general the former is much greater than the latter. Therefore the errors made by the pyramid builders in the positioning of these monuments, supposing they really wanted to represent the asterism of the Orion Belt, fall within the uncertainty range of the stellar positions known at that time. In other words, according to Bauval e Gilbert (1994), *the positions of the main pyramids of the Giza plain correspond to those of the Orion Belt stars within the error margin of the astrometric measurements of that epoch.*

It is interesting to recall that, to overlap the Orion Belt stars on the Giza pyramids, it is first of all necessary to rotate of 180° the celestial map, so that the northernmost star (Mintaka) corresponds to the southernmost pyramid (Menkhaure). As suggested by Bauval (2006), this apparent reversal of the North-South axis does not constitute a problem at all, since it could be simply due to the fact that ancient Egyptians probably drew their geographic maps with South "at the top". Obviously this choice is opposite to that adopted by the cartographers of the XVII century who decided to put North on the top of their maps, a convention that we too continue to make today. In any case there is no objective reason to put necessarily North at the top of the geographic charts. All the orientations are possible; it is only a matter of conventions. For example in the Medieval map known as the Hereford Mappa Mundi (dating to XIII century) East is at the top. According to Bauval (2006), for ancient Egyptians it was more logical to put South, and not North, on the top of their maps. South was "up" since the Nile river flows down from South and since the Sun culminates (reaches its highest point in the sky) exactly in the South at midday. Actually, the Egyptians called (and we still call) the southern part of their country as "Upper Egypt" and the northern one as "Lower Egypt". Therefore it would not be surprising that Mintaka, the upper star of the Orion Belt (at its culmination), was associated with the Menkhaure pyramid, the "upper" one in the hypothetical topographic maps of that time.

It is also interesting the reason of the additional anti-clockwise rotation of 15.3°, required to overlap the Orion Belt stars on the corresponding pyramids. Actually, such a rotation is necessary since the "axis of the Pyramids" (i.e. the straight line that best fits the positions of the centers, or vertexes, of the three Giza pyramids) is tilted of about 38° with respect to the North-South direction, while the "axis of the Orion Belt" (the straight line that best fits the positions of the stars of the asterism) is tilted of about 53° with respect to the celestial North-South direction. Therefore it is necessary a 15.3° anticlockwise rotation of the star map in order to superimpose the two axes within the errors.

The problem of the different inclinations of the axes of the pyramids and of the Orion Belt with respect to the corresponding North-South direction (terrestrial, in the first case, celestial in the second one) is an important question that will be discussed in detail in the next section. Here it is important to note that another test has been done in order to find around Giza other examples of pyramids located in such a way to reproduce on the ground the position of other stars, besides the Orion Belt ones. In particular the same star chart previously used for the fit shown in Fig. 1, but extended up to 40° around the Orion Belt (in order to include various luminous stars in Orion and other constellations) has been superimposed upon a map of the pyramid sites; then, the chart has been rescaled and rotated until the Orion Belt stars matched exactly with the three pyramids at Giza. At this point other coincidences star-pyramid have been searched for. This search has been completely unsuccessful: apart from the Orion Belt stars and the Giza pyramids, no other correspondence exists between stars and pyramids neither in Orion nor in other constellations; in particular, Saiph is more than 22° far from the celestial point corresponding to the Abu Roash pyramid, while Bellatrix is about 12° far from the celestial point corresponding to the Zawiyet el-Aryan pyramid. In other words, contrary to the what reported by Bauval e Gilbert (1994), the match stars-pyramids is completely unsatisfactory, unless restricted to the case Orion Belt stars – Giza pyramids. The present result is in perfect agreement with those obtained by other authors (Legon, 1995; Orcutt, 2000) and argues against the existence of a hypothetical unified "master plan" according to which five pyramids in the Giza area were laid out on the ground to reproduce the pattern of the stars in the Orion constellation (Bauval e Gilbert, 1994). In any case, the simplest version of the OCT (correspondence limited only to Orion Belt stars and Giza pyramids) is not ruled out by this test.

### 3. Orientation of the pyramid axis

The inclination of the axis of the Pyramids does not seem to be casual. Actually the south-eastern corners of the three Giza pyramids are aligned in good approximation towards the great solar temple of Heliopolis (Lehner, 1985a,b; Goedicke, 2001; Magli, 2009a,b). More precisely the straight line connecting the south-eastern corners of the two extreme pyramids (those of Menkaure and Khufu) passes about 12 m away from the corresponding corner of the central (Khafre) pyramid; however this discrepancy is less than 2% of the distance (637 m) between the corners of the extreme pyramids. It is then likely that the disposition of the main pyramids on the Giza plain was dictated by religious considerations.

However, it remains to verify whether this orientation was also due to astronomical motivations more or less linked to the Orion constellation. In this respect it is worthwhile to report an often cited hypothesis proposed by Bauval (2006) regarding the coincidence between the angle formed with the terrestrial North-South direction by the straight line connecting the centers of the pyramids of Menkaure and Khufu (hereinafter called "$P_1P_3$ axis") and the angle formed with the celestial North-South direction by the straight line connecting Mintaka and Alnitak (hereinafter "$S_1S_3$ axis"). While at Giza the angle between the $P_1P_3$ axis and the North-South direction is 37.8° (Petrie, 1883), on the celestial sphere the angle between the $S_1S_3$ axis and the North-South direction is 53.1° today and it was even larger (73.9°) at the time of pyramids, due to the precessional motion of Earth's axis, also known as the precession of the equinoxes (see Fig. 2). Actually, this effect, due to the combined gravitational actions of the Moon and the Sun on the Earth, gives rise to a slow cyclical change in the position of stars, as measured in the equatorial coordinate system, with a period of about 26000 years. According to Bauval (2006) it would be necessary to go back to 11500 BC in order to have the $S_1S_3$ axis tilted exactly by 37.8° with respect to the celestial North-South direction: in this case the arrangement of the main pyramids on the Giza plain would precisely reflect that of the stars of Orion Belt in the sky. In particular, Bauval (2006) claims that the Giza pyramids were built during the Fourth Dynasty, but they were located in such a way to reproduce on the ground the position of the Orion Belt stars on the celestial sphere well 9000 years before (at the epoch of the *Zep Tepi*, the Beginning of Time, according to the Egyptian creation myth). This ancient position of the asterism would have been calculated by the Egyptian astronomer-priests on the basis of their knowledge of the phenomenon of the precession of the equinoxes or would have been reconstructed, using old star maps dating back to 11500 BC (Bauval, 2006).

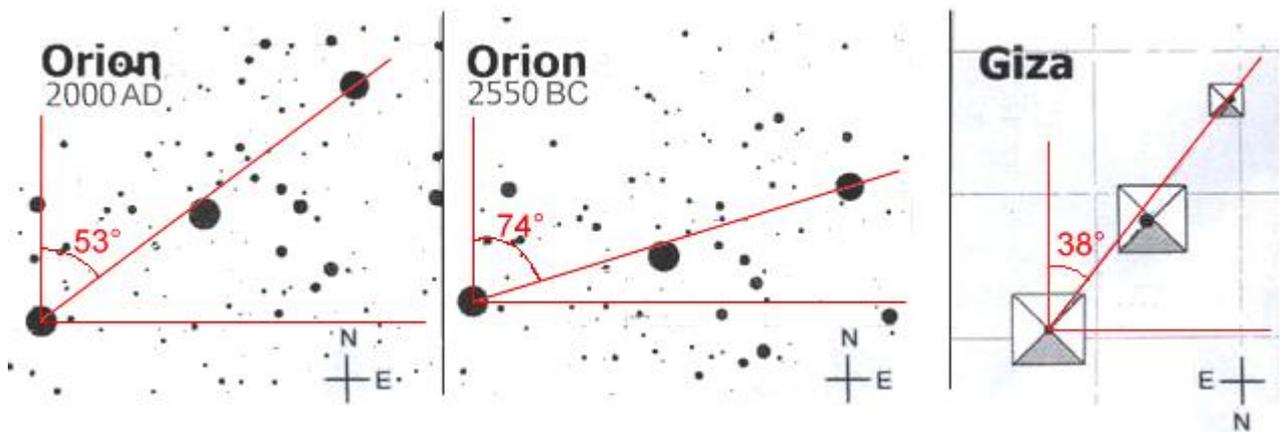

**Figure 2** – Position of the Orion Belt on the celestial sphere in two different epochs (today and 4600 years ago) compared with the position of the Giza pyramids; note that North is up in the two star charts and down in the topographic map of Giza. The changing aspect of the Orion Belt is due to the precession of the equinoxes.

However, also admitting that the ancient Egyptians were aware of the precession of the equinoxes (a fact which has never been definitively attested), it is difficult to believe that they had such a knowledge of the precessional motion to be able to calculate its effects on the position of the asterism back in time for many thousands of years (Tedder, 2006; Magli and Belmonte, 2009). This also considering the large uncertainties that even today affect the results of the commercial routines of precessional computation when they are used for so remote epochs (Fairall, 1999). On the other hand, it is also difficult to think that, since 11500 BC (that is in the heart of prehistory), the ancient Egyptians were able to draw detailed stellar maps passed on from generation to generation up to the Fourth Dynasty. In conclusion, also this aspect of the OCT, if not wrong, is at least very unlikely.

In any case, as suggested by Tedder (2006) it is not necessary to go so far back in time to discover a satisfactory coincidence between the arrangement of the Giza pyramids and that of the stars of the Orion Belt, since just during the Fourth Dynasty it is possible to find an astronomical configuration very interesting in this respect. In particular, a winter day around 2550 BC, an observer located at Giza could see Alnitak vertically aligned with Saiph at an azimuth of 142.2°. In such a configuration the $S_1S_3$ axis was tilted with respect to the vertical of an angle close to that formed by the $P_1P_3$ axis with the direction North-South (Tedder, 2006). In other words, when the vertical alignment Alnitak-Saiph occurred, the arrangement of the stars of the Orion Belt in the sky mirrored that of the main pyramids in the Giza plain (with the vertical direction that played on the celestial vault the same role of the local meridian on the ground – see Tedder (2006) for details).

It is important to stress the importance (not only astronomical) of the direction of observation of this alignment, since all the main groups of pyramids were located along a straight line (or on the sides of it) oriented just to 52.2° South of East: the northernmost end of this line coincided with the pyramid of Djedefra at Abu Roash, while the southernmost one was located at the pyramid of Userkaf (the founder of the Fifth Dynasty) near Saqqara (see Tedder (2006) for details).

Another possible astronomical motivation for the orientation of the $P_1P_3$ axis could be linked to the correspondence between the position of the Giza pyramids with respect to the Nile and that of the Orion Belt stars with respect to the Milky Way. In this context (see Fig. 3), if one indicates with $P_1$ and $P_3$ the vertices of Menkaure and Khufu pyramids, respectively, with $N_1$ and $N_2$ the intersections of the axis $P_1P_3$ with the western and the eastern bank of the Nile, respectively, and finally with N the center of the Nile along the same axis, then, using the tool ruler of the software Google Earth (downloadable from the site http://www.google.com/intl/it/earth/index.html), it is easy to see that:

$R = P_1N / P_1P_3 = (P_1N_1 + P_1N_2) / 2 P_1P_3 = 16.89 \pm 0.01$

(where the uncertainty has been determined evaluating the accuracy of the tool ruler in measuring known distances). In other words the distance, along the axis $P_1P_3$, between the Menkaure pyramid and the Nile is equal to 16.9 times the distance between the two pyramids at the ends of the alignment of Giza.

Now, assuming a correspondence between pyramids and stars, one could search for the point of the celestial vault which is the equivalent in the sky of the center of the Nile (as defined above), that is the point lying on the extension of the segment $S_1S_3$ connecting Mintaka with Alnitak, at a distance from the former equal to 16.9 times the angular separation between the two stars. This point is labeled with C on the star map shown in Fig. 4: as it can be seen it is placed in the hearth of the innermost region of the Milky Way visible by naked eyed, very close to the central point $C_0$ of this region along the $S_1S_3$ axis.

Similarly to the case of the pyramids, $C_0$ can be determined, as shown in Fig. 4, if one indicates with $S_1$ and $S_3$ the positions of Mintaka and Alnitak, respectively, with $C_1$ the starting point (along the $S_1S_3$ axis) of the visible and the innermost part of the Milky Way and with $C_2$ the ending point of this region (along the same axis); evidently $C_0$ is the mean point of the segment $C_1C_2$. For example, in the case of the map shown in Fig. 4, one finds that:

$R' = S_1C_0 / S_1S_3 = 16.1$.

This means that, if we move on the extension of the segment connecting Mintaka to Alnitak, then the angular distance between the former and the center of the Milky Way (defined above) is 16.1 times the angular separation of the two stars at the ends of the asterism. The two ratios R and R' differ of about 4.7%.

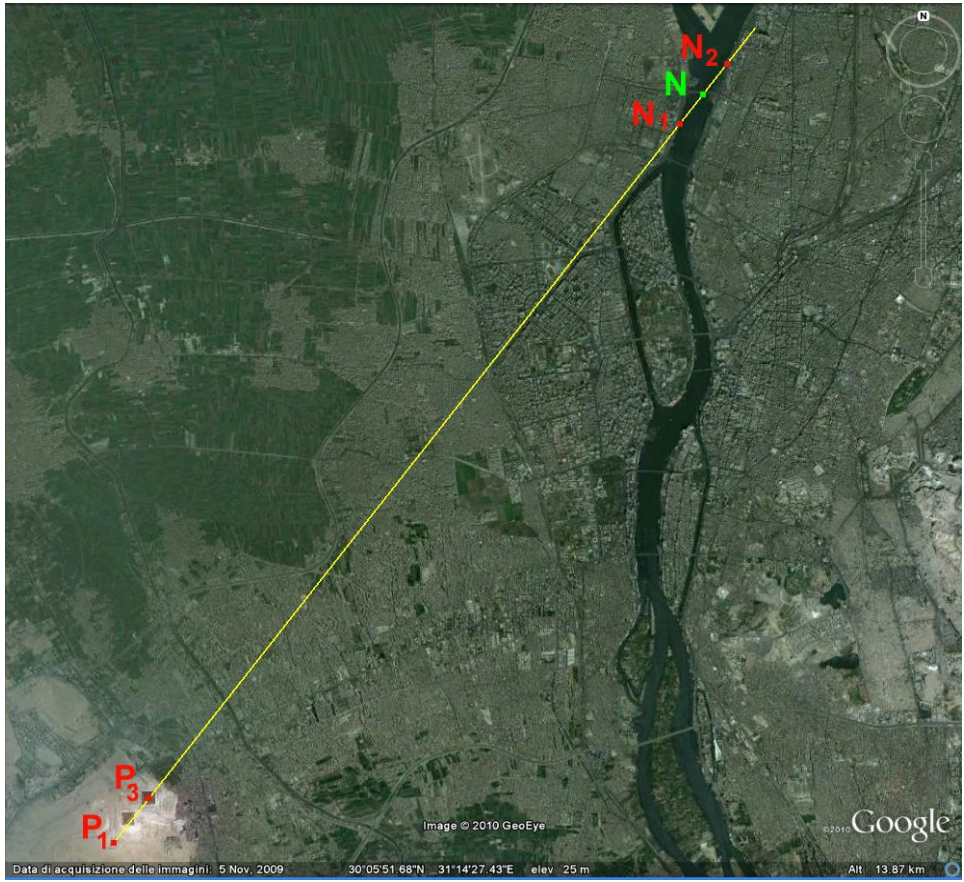

**Figure 3** – Topographic map showing the position of the Giza pyramids with respect to the Nile river. $P_1$ and $P_3$ indicate the pyramids of Menkaure and Khufu, respectively, while N represents the point, located along the axis $P_1P_3$, which is at the center of the river. Points $N_1$ and $N_2$ are defined in the text (Credit © 2010 Google Earth /Image © 2010 GeoEye).

It is important to note that, even if this discrepancy between C and $C_0$ is not negligible, it does not seem significant due to the following reasons. Even for a trained eye and in perfect visibility conditions, the Milky Way is barely distinguishable and therefore it is difficult to detect the centre of this faint belt, also because of its inhomogeneous brightness distribution. Furthermore the position of $C_0$ in the stellar maps critically depends on the way in which the boundaries of the Milky Way are drawn: since in the sky sharp border lines are lacking, they are always plotted in a somewhat arbitrary way. Actually, using different star charts (download from Internet or generated by various astronomical software) the resulting position of $C_0$ varies considerably, implying a conspicuous uncertainty in the value of the R' ratio. In particular, from the various analyzed star maps I obtained R' = 16.5 ± 0.9. Comparing this result with the value R = 16.89 ± 0.01, which allows to locate the point C, it is possible to conclude that the celestial counterpart of the center of Nile river falls well within the uncertainty range in the position of the center of the Milky Way, as reported by the various star charts.

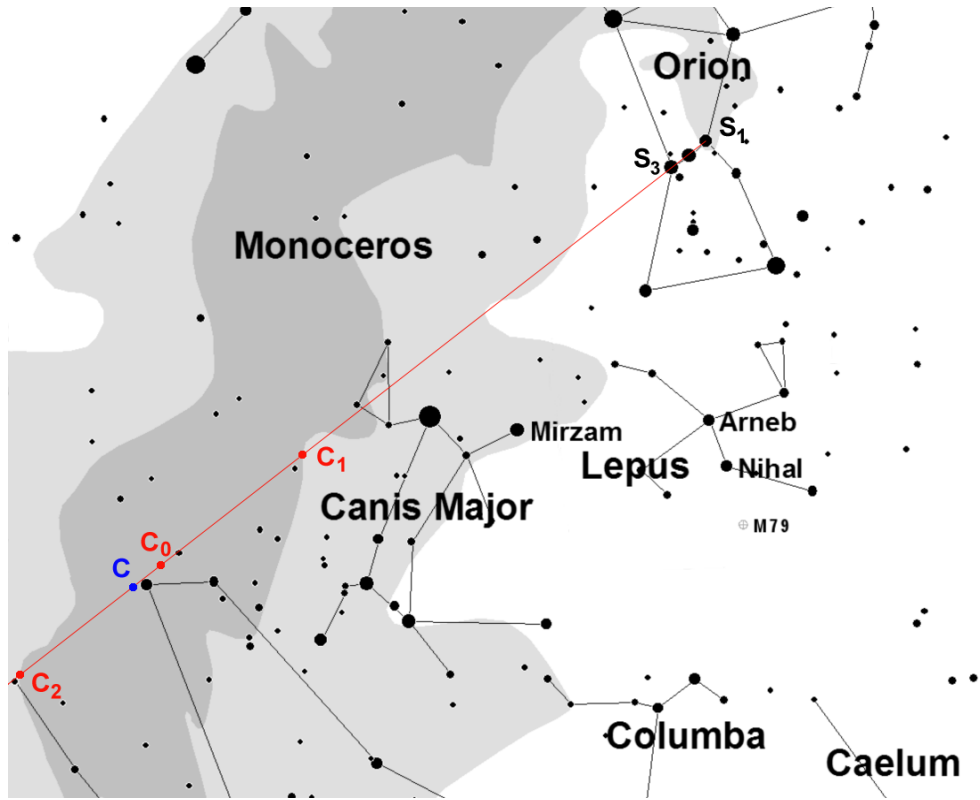

**Figure 4** – Star map (from the web site http://www.desertexposure.com/200902/200902_starry.php) showing the position of the Orion Belt stars with respect to the Milky Way (gray areas). $S_1$ and $S_3$ are Mintaka and Alnitak, respectively, $C_0$ represents the point on the extension of the $S_1 S_3$ segment that is at the center of the innermost region of the Milky Way which is visible by naked eye (dark gray area), while C is the celestial counterpart of the center of Nile river, as defined in the text.

In other words, in the scale that would have been chosen by the pyramid builders to reproduce the Orion Belt on the ground, the linear distance between the Menkaure pyramid and the Nile, along the straight line connecting the two extreme pyramids of Giza, is very close (practically coincides within the above discussed uncertainties) with the angular distance between Mintaka and the central point of the Milky Way along the straight line connecting the two extreme stars of the asterism.

In summary, a quantitative correspondence would exist between the position of the pyramids of Giza with respect to the Nile and that of the stars of the Orion Belt with respect to the Milky Way. Obviously this correspondence is hypothetical, also considering the variations in the course of Nile river that could have occurred in the last 5000 years. However, such a coincidence is noteworthy because, according to various authors (Lamy, 1981; Bauval, 1989; Lehner, 1997), the ancient Egyptians identified the Nile with the Milky Way, in the sense that for them this faint belt of stars represented some sort of river in the sky, the celestial counterpart of their river. All the burials occurred on the western bank of the Nile that, along with the pyramid fields, symbolized the Orion region on the fringes of the Milky Way. In the Egyptian funeral rituals the transport of the remains of the dead across the Nile for the burial was in some way linked to the journey of the soul coming into the Osiris kingdom across the celestial Nile, the Milky Way. The latter was, therefore, a sort of Styx, the river of the Underworld, that the dead had to cross in order to reach the next life.

## 4. Comparison between the dimensions of the Giza pyramids and the magnitudes of the stars of the Orion Belt

In order to weigh up in detail the OCT, it is necessary to verify if the dimension of the pyramids is correlated with the brightness of the Orion Belt stars. In this respect one can consider two main dimensions: height and volume. The former is the more evident one, since it is directly evaluable by only one measurement, even if the volume obviously provides the more precise and complete information, also because, contrary to height (which is always relative to a reference level – see below), the volume gives an absolute evaluation of the pyramid dimension, directly comparable with that of the other ones.

The volume of the pyramids has been evaluated using the original values of the height and of the side of base; these quantities are listed in Table 1, where the visual magnitude of the corresponding Orion Belt stars is also reported. The latter is a measure of the apparent brightness of the object and it is defined by the Pogson formula:

$$m = a \log F + m_0, \tag{1}$$

where $F$ is the flux (energy per unit time and area) of the visible radiation received by the observer, $a$ is constant equal to $-2.5$, while $m_0$ is another constant conventionally chosen by assigning to a reference-star an *a priori* fixed magnitude (the original choice was to use as a reference the Polar Star, assigning to it a visual magnitude $m = 2$). By fixing in this way the two constants $a$ and $m_0$, Pogson made the modern star magnitude scale consistent with the ancient photometric classification of the stars performed in the II Century BC by Hipparchus. For the same historical reason, magnitude decreases with increasing flux (i.e. the brightest objects have the lowest magnitudes).

Since both Mintaka and Alnilam are variable stars, for them I used the average magnitudes reported, along with the magnitude of Alnitak (which is constant in time), in the recent *Catalogo de Magnitudes Aparentes* (Catalogue of Apparent Magnitudes – Otero, 2014). This catalogue, which will be completed shortly, provides for the first time combined information about the variability and the molteplicity of all the stars with a visual magnitude less than 5. Note that the magnitude of the three stars of the Belt are in good agreement with those reported in the work by Hardie et al. (1964) which is the most recent peer-reviewed paper on the subject.

Table 1 shows that, as suggested by the OCT, the fainter star of the Belt is actually associated with the smallest and lowest pyramid of Giza (Menkaure); however, the problem exists that the brightest star (Alnilam) is not associated with the largest and tallest pyramid (Khufu). The latter is instead associated with the star with intermediate brightness (Alnitak). In other words there is no correlation between the magnitude of the stars of the Belt and the height or the volume of the corresponding pyramids. This happens, however, if one considers the intrinsic height of the pyramids, that is the height of their vertex with respect to their base level.

Things changes considerably if one considers, instead of the intrinsic height of each pyramid, the one that can be called "apparent" height, evaluated with respect to a common reference level (the same for the three pyramids), such as the sea level or the lowest among the base levels of the three pyramids. This quantity is exactly the one we take into consideration, for example, when we compare the heights of the mountains. Actually, the height that matters for an observer who looks at the pyramids from a distance is the apparent height, that does not coincide with the intrinsic one, since the pyramids of Menkaure and Khafre are built on a plain at about 70 m on sea level, while the base of Khufu pyramid is located about 10 m lower down (Lehner, 1985a). This can be easily seen from the topographic data provided by the software Google Earth.

**Table 1** – Original values of the side of base (*l*), the height (*h*) and the volume (*V*) of the three Giza pyramids, compared with the visual magnitude (*m*) of the three stars of the Orion Belt (Otero, 2014). It is also reported the apparent height (*h'*) of each pyramid, that is the height of the vertex with respect to the base level of Khufu pyramid.

| Pyramid | *l* (m) | *h* (m) | *h'* (m) | *V* (m$^3$) | Star | *m* |
|---|---|---|---|---|---|---|
| Menkaure (Mykerinos) | 105 | 66 | 76 | $2.43 \times 10^5$ | Mintaka (δ Orionis) | 2.23 |
| Khafre (Chephren) | 215 | 144 | 154 | $2.22 \times 10^6$ | Alnilam (ε Orionis) | 1.69 |
| Khufu (Cheops) | 231 | 147 | 147 | $2.61 \times 10^6$ | Alnitak (ζ Orionis) | 1.76 |

NOTE: The dimensions of the pyramids of Menkaure, Khafre and Khufu have been taken respectively from the sites:
http://www.nationmaster.com/encyclopedia/Menkaure's-Pyramid,
http://www.nationmaster.com/encyclopedia/Pyramid-of-Khafre,
http://www.nationmaster.com/encyclopedia/Great-Pyramid-of-Giza.

A plot of the visual magnitude *m* of the Orion Belt stars versus the apparent height *h'* of the corresponding pyramids with respect to the base level of the Khufu pyramid shows a significant anticorrelation between the two quantities (see Fig. 5). Such an anticorrelation means that equal increases in *h'* correspond to equal decreases in *m* and therefore to equal increments in the apparent brightness of the stars. The ancient Egyptian well knew this kind of geometric-mathematical relationship that they used many times when, for example, they planned and carried out architectural structures with constant slope (where equal horizontal displacements correspond to equal vertical displacements), such as the same pyramids and the shafts and the corridors inside them.

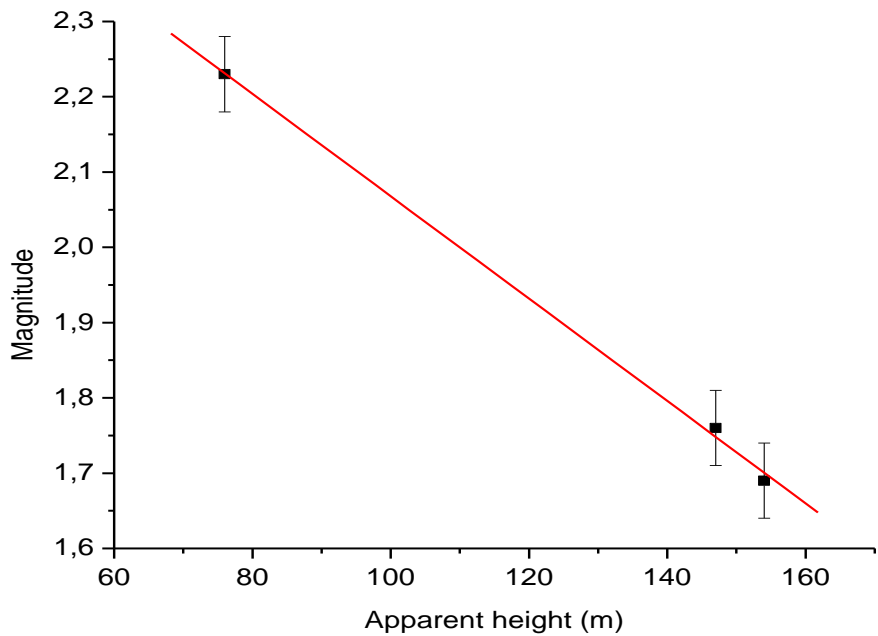

**Figure 5** – Straight line of linear regression between the apparent height (*h'*) of the pyramids and the visual magnitude (*m*) of the corresponding stars of the Orion Belt (the error bar is due to the typical uncertainty of the evaluation by naked eye, equal to ±0.05 – Otero and Moon, 2006). The linear anticorrelation between *h'* and *m* implies a linear correlation between *h'* and the apparent brightness of the Orion Belt stars, that is a linear grow of the latter with the apparent height of the corresponding pyramids.

It is worthwhile to note the correlation coefficient (Taylor, 1982) associated to the data shown in Fig. 5 is equal to − 0.9993, and in principle this value would imply a probability of only 3% or lesser that the anticorrelation between *h'* and *m* could be due to sheer chance. Even so, since one cannot carry out a statistical examination of only three data, one cannot exclude that the trend shown in Fig. 5 is fortuitous and not wanted by the pyramid builders. In any case this result removes one of the most serious objections to the OCT, that is the claimed lack of correlation between the size of Giza pyramids and the brightness of the Orion Belt stars, while it is in agreement with what reported by Bauval and Gilbert (1994), who in their book qualitatively state that *"not only did the layout of the pyramids match the stars with uncanny precision but the intensity of the stars, shown by their apparent size, corresponded with the Giza group…."*.

## 5. Astronomical verification of the OCT

The correlation found in the previous section has to be subjected to two important astronomical/ astrophysical tests in order to evaluate its reliability. The first one consists in checking whether the magnitude of Alnilam is such that, when observed by naked eye, this star is appreciably brighter than Alnitak. This is important because otherwise the two main pyramids should have the same apparent height, contrary to what we observe at Giza.

In this regard, it is worthwhile to note that today the amateur astronomers who deal with naked eye observations of variable stars (variabilists) use appropriate comparison techniques that in general allow to appreciate small differences in the stellar magnitude (Otero and Moon, 2006). This is particularly true in the case of Alnitak and Alnilam, because of the closeness of the two objects, their comparable brightness and their very similar color. Even the beginners, when asked to give their impressions, assert they are able to detect this slight difference (Otero, 2010). Then, if we assume that the Egyptian priest-astronomers, for centuries skilled in naked eye observations, acquired the same techniques adopted by the modern variabilists, then it is not unreasonable to think that they too were able to appreciate the different magnitude of the two stars, also considering the dry and smogless atmosphere of the ancient Egypt.

Another thing to be assessed is whether in the past the magnitudes of the Orion Belt stars were different from the present ones, implying that the above discussed correlation could not be valid at that time. Actually Mintaka, Alnitak and Alnilam (especially the last two) are blue supergiants that could be in an advanced evolutionary phase, and, in principle, could experience conspicuous and fast changes in its luminosity. However, as thoroughly discussed by Orofino (2011), the models of stellar evolution suggest that the three stars of the Orion Belt are evolved but stationary objects, in the sense that their visual magnitude has remained practically the same, at least in the last 10 thousand years; for this reason the correlation found in the previous section was still valid when the pyramids were built.

In summary, it is possible to state that *the OCT is consistent with what expected for Alnitak, Alnilam and Mintaka on the basis of the stellar photometry and stellar evolution*.

**6. Discussion and conclusions**

The results found in the previous sections can be summarized as follows:

a) the relative positions of the three Giza pyramids coincide, within the uncertainties of the naked-eye astrometric measurements, with the relative positions of the three stars of the Orion Belt;

b) in the scale that would have been chosen by the pyramid builders to reproduce the Orion Belt on the ground, the linear distance between the Menkhaure pyramid and the Nile, along the straight line connecting the two extreme pyramids of Giza, practically matches the angular distance between Mintaka and the central point of the Milky Way along the straight line connecting the two extreme stars of the asterism;

c) the visual magnitude of the stars of the Belt is presently correlated with the height of the corresponding pyramids evaluated with respect to a common reference level (i.e. the base level of the Khufu pyramid).

Since the star evolution models suggest that the magnitudes of all the three objects of the Belt at the time of the pyramids were substantially equal to the present ones, the above found correlation was still valid at that epoch (Orofino, 2011).

In the light of the previous results one can conclude that the OCT, in its simplest version (see sect. 2), is not incompatible with what expected for the stars of the Orion Belt on the basis of naked-eye astrometry and photometry, as well as of the stellar evolution theory. Therefore, there are no astronomical/astrophysical arguments to reject the hypothesis that the main Giza pyramids would represent the monumental reproduction on the ground of the Orion Belt, with the Nile river representing the Milky Way (Bauval and Gilbert, 1994; Bauval, 2006). According to Bauval (2006), all this would fit into the framework of the belief, widespread among the ancient Egyptians, of a correspondence between Heaven and Earth summarized by the hermetic dictum "as above, so below".

An important question concerns the correlation found and discussed in section 4. Actually, even if one assumes that such a correlation is intentional, it is very likely that the Egyptian project was the realization of a simple qualitative correspondence between the apparent height of the pyramids and the visual magnitude (that is the apparent brightness) of the relative stars. According to such a project, the central pyramid at Giza had to be slightly higher (in terms of apparent height) than one of the extreme pyramids of the group and much higher than the other one, in order to mirror the fact that the central star of the Belt is slightly brighter than one of the extreme stars of the asterism and much more luminous than the other one. However, if we admit that the correlation between the apparent height of the pyramids and the magnitude of the stars is not only intentional and but also quantitatively valid, then we could conclude that the ancient Egyptians were able to evaluate the stellar magnitudes with a good level of accuracy; but this remains only a simple hypothesis, since there are no textual proofs for such an observational skill of the priest-astronomer of the Old Kingdom.

In any case, it cannot be excluded that the observation of decans and the study of the lunar phases, of which sure traces exist in the archaeological finds dating back to the Medium Kingdom, have actually their roots in much older astronomical practices. This hypothesis is supported by some archaeological discoveries made at Nabta Playa, a locality in the Nubian Desert 800 km south of Cairo (Malville et al., 1998, 2007). Here, according to some authors (Shild e Wendorf, 2004), a semi-nomadic population lived in the sixth millennium BC, when the climate was milder than today. Later, such a population, due to the worsening of the climatic conditions, migrated to the fertile Nile valley where originated the Egyptian civilization that we know today (Malville et al., 1998). The archaeological finds of Nabta Playa concern a small stone circle, measuring roughly 4 m in diameter, that was probably an astronomical calendar, as well as six megalithic alignments, extending for up to a mile, that would point in the directions where some of the brightest stars of the sky, in particular Sirius, Alnilam and the other stars of the Orion Belt, raised between 4600 and 4200 BC (Malville et al., 2007). The presence of these stars, so important for the Egyptian astronomy and mythology (see Section 1), could not be casual: the great interest of the Egyptians for the celestial phenomena in general, and for such stars in particular, would have been inherited from their putative Nubian ancestors.

According to such point of view, the Egyptian astronomy could not be started in the epoch of the first purely astronomic papyri (Middle Kingdom), but may have had much more ancient origins, coeval or even older than those of astronomy in Mesopotamia, universally acknowledged as the cradle of this science. In conclusion, the paradigm of a very primitive Egyptian astronomy during the Old Kingdom is probably not as obvious as commonly believed by the most of the Egyptologists.

**Acknowledgments**

The author warmly thanks Sebastian Otero and Chris Tedder for the very useful discussions on various subjects treated in the text. Giulio Magli is also thanked for his comments and suggestions that helped to improve the quality of the manuscript.